\documentclass[conference]{IEEEtran}
\IEEEoverridecommandlockouts

\usepackage{cite}
\usepackage{amsmath,amssymb,amsfonts}
\usepackage{algorithmic}
\usepackage{graphicx}
\usepackage{textcomp}
\usepackage{xcolor}
\def\BibTeX{{\rm B\kern-.05em{\sc i\kern-.025em b}\kern-.08em
    T\kern-.1667em\lower.7ex\hbox{E}\kern-.125emX}}
\begin{document}

\title{A hybrid approach for improving U-Net variants in medical image segmentation\\}

\author{\IEEEauthorblockN{Aitik Gupta}
\IEEEauthorblockA{\textit{Student, Information Technology} \\
\textit{ABV-IIITM, Gwalior}\\
Madhya Pradesh, India - 474015 \\
imt\_2018010@iiitm.ac.in}
\and
\IEEEauthorblockN{Dr. Joydip Dhar}
\IEEEauthorblockA{\textit{Department of Applied Sciences} \\
\textit{ABV-IIITM, Gwalior}\\
Madhya Pradesh, India - 474015 \\
jdhar@iiitm.ac.in}
}

\maketitle

\begin{abstract}
Medical image segmentation is vital to the area of medical imaging because it enables professionals to more accurately examine and understand the information offered by different imaging modalities. The technique of splitting a medical image into various segments or regions of interest is known as medical image segmentation. The segmented images that are produced can be used for many different things, including diagnosis, surgery planning, and therapy evaluation.

In initial phase of research, major focus has been given to review existing deep-learning approaches, including researches like MultiResUNet, Attention U-Net, classical U-Net, and other variants. The attention feature vectors or maps dynamically add important weights to critical information, and most of these variants use these to increase accuracy, but the network parameter requirements are somewhat more stringent. They face certain problems such as overfitting, as their number of trainable parameters is very high, and so is their inference time.

Therefore, the aim of this research is to reduce the network parameter requirements using depthwise separable convolutions, while maintaining performance over some medical image segmentation tasks such as skin lesion segmentation using attention system and residual connections.
\end{abstract}

\section{Introduction}
\subsection{Medical Image Segmentation}\label{Chapter_1_Motivation}
\noindent Medical image segmentation is the process of dividing a digital medical image into multiple segments, or regions, based on certain criteria. This is often done to highlight specific structures or abnormalities in the image, and can be useful for diagnosis and treatment planning. There are many different techniques and algorithms used for medical image segmentation, including thresholding, clustering, and deep learning methods.

Traditionally, medical images are segmented by manually outlining the desired targets, which is a labour-intensive and specialized task requiring physician expertise. In the past, many automated methods based on image morphology have been proposed, including edge detection, area detection, and template matching. These methods aim to automate the segmentation process and reduce the reliance on manual labour.



\subsection{Problem Statement}
\noindent As deep learning approaches are becoming prominent in medical image segmentation tasks, there is a need for optimized approaches which are quick for inferencing, as well as reasonably accurate.

Therefore, the problem statement of the research is to optimize the number of trainable parameters of recent approaches \cite{9246575} \cite{IBTEHAZ202074}, while maintaining the performance on medical image segmentation tasks such as skin lesion segmentation \cite{Tschandl2018} using depthwise separable convolutions \cite{xceptionpaper} and attention mechanism inspired from \cite{2018arXiv180403999O}.

\subsection{Traditional Methods}
\subsubsection{Thresholding}\label{thresholdingmethod}
One common approach to medical image segmentation is thresholding, which involves dividing the image into two or more regions based on a threshold value. For example, a simple thresholding algorithm might divide an Magnetic Resource Imaging (MRI) scan into two regions: one region representing the brain tissue and another representing the surrounding fluid. Thresholding algorithms can be useful for identifying structures of interest in an image, but they can also be sensitive to noise and other variations in the image.

Thresholding is a common technique used in medical image segmentation to convert a continuous-valued image into a binary image, where each pixel is either foreground (object of interest) or background. This is done by selecting a threshold value, and then assigning all pixels with intensities above the threshold to the foreground, and all pixels with intensities below the threshold to the background.

There are many different ways to choose the threshold value for thresholding, and the appropriate method will depend on the specific application and the characteristics of the image. Some common methods for choosing the threshold include:

\begin{figure}[h]
    \centering \includegraphics[width=0.5\textwidth]{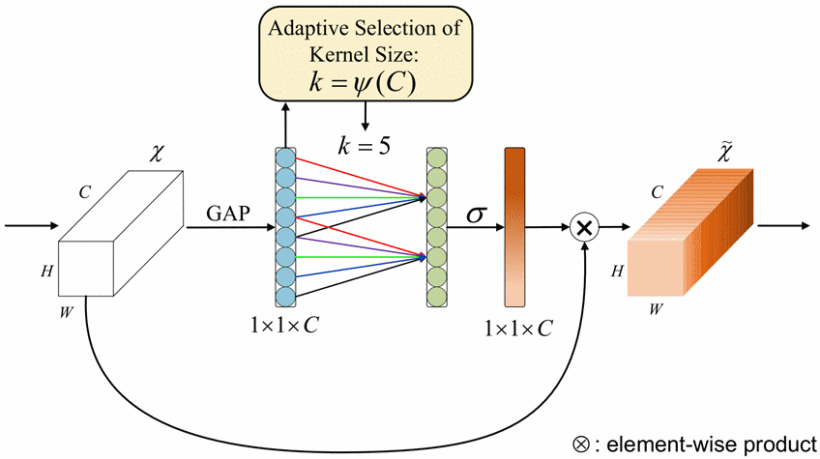}
    \caption{X-Element-wise product}
\end{figure}

\paragraph{Global thresholding}
This involves selecting a single threshold value for the entire image, based on the overall intensity distribution of the image. This can be useful for images with a clear separation between foreground and background pixels.

\paragraph{Adaptive thresholding}
This involves selecting a different threshold value for each pixel, based on the local intensity values in the image. This can be useful for images with varying foreground and background intensities, or for images with non-uniform illumination.

\paragraph{Otsu's method}
This is a popular global thresholding method \cite{GOH2018298} that automatically determines the optimal threshold value based on the intensity histogram of the image. It assumes that the image contains two classes of pixels (foreground and background), and it tries to find the threshold that minimizes the intra-class variance.

In general, thresholding is a simple but powerful technique for medical image segmentation, and it is widely used in many different applications. It can be particularly useful for images with clear foreground and background intensity distributions, but it may not be suitable for all types of images or segmentation tasks.

\subsubsection{Clustering}\label{clusteringmethod}
Another popular approach to medical image segmentation is clustering, which involves dividing the image into multiple regions based on the similarity of the pixel values within each region. Clustering algorithms can be useful for identifying structures with similar characteristics, such as different types of tissue in an MRI scan. However, clustering algorithms can be sensitive to the choice of similarity measure and the number of clusters used.

\paragraph{K-Means clustering}
In the context of medical image segmentation, the goal of the algorithm would be to group together similar pixels in an image. This can be useful for identifying distinct objects or tissue types in a medical image, such as tumors or healthy tissue.

To apply K-means clustering to a medical image, the algorithm first needs to be trained on a set of labeled images, where the pixels in each image have been manually grouped into the desired clusters. Once the algorithm has been trained, it can then be applied to new images to automatically segment the pixels into the learned clusters.

In general, K-means type of clustering works by dividing the pixels in an image into a specified number of clusters, called "k" clusters. The algorithm works by iteratively assigning each pixel to the cluster with the nearest mean, or average value. After each assignment, the mean value of each cluster is recalculated based on the pixels that are currently assigned to that cluster. This process continues until the cluster assignments stop changing, at which point the algorithm has converged and the final cluster assignments are returned.

One advantage of using K-means clustering for medical image segmentation is that it is relatively fast and easy to implement. It can also produce good results when the clusters in an image are well-defined and well-separated from each other.

The algorithm can be sensitive to the initial cluster assignments, and it may not always produce the best possible segmentation of an image.
\subsection{Deep Learning Methods}
Deep learning methods have become increasingly important for medical image segmentation in recent years. This is because deep learning algorithms are able to automatically learn rich, hierarchical features from large amounts of data, which allows them to achieve superior results on a variety of tasks. In the case of medical image segmentation, deep learning algorithms are able to learn the complex patterns and structures present in medical images, which allows them to segment different structures and objects within the images accurately.

\subsubsection{Advantages of Deep Learning methods}
\noindent The use of deep learning methods for medical image segmentation has several advantages.

\begin{itemize}
    \item First, deep learning algorithms can learn to segment images without requiring explicit, hand-crafted features, which makes them more flexible and adaptable to a wider range of tasks.
    \item Second, deep learning algorithms can learn from large, annotated datasets, which allows them to improve their performance over time as more data becomes available.
    \item Third, deep learning algorithms can operate on high-resolution images, which allows them to capture fine-grained details and produce high-quality segmentation maps.
\end{itemize}

Overall, the use of deep learning methods for medical image segmentation has the potential to greatly improve the accuracy and efficiency of medical image analysis, which can lead to better diagnosis and treatment of medical conditions. These methods have been shown to be effective for a wide range of medical image segmentation tasks \cite{9356353}, including segmenting organs \cite{FU2021107}, tumors \cite{braintumorsegmentation}, and other structures in medical images.

\subsubsection{Convolutional Neural Networks}
Convolutional Neural Networks (CNNs) are a type of deep learning algorithm that is specifically designed to process data with a grid-like structure, such as images. A CNN is composed of a series of convolutional layers, which apply a set of learnable filters to the input image to extract features at different scales and locations. These features are then passed through a series of fully connected layers, which combine them to produce a segmentation map.

\paragraph{Kernels}
The key idea behind convolutional layers is the use of filters, which are small, learnable kernels that are applied to the input data. Each filter slides across the input data, computing a dot product between the filter weights and the input data at each location. This process produces a feature map, which encodes the presence of certain patterns or features within the input data.

\par Convolutional layers can be stacked to form a CNN, which allows the network to learn hierarchical, multi-scale features from the input data. For example, a CNN can learn to detect edges in an image in the first convolutional layer, then learn to combine these edges to detect shapes in the second convolutional layer, and finally learn to combine these shapes to recognize objects in the third convolutional layer.

\paragraph{Pooling Layers}
In addition to convolutional layers, CNNs also include pooling layers, which are used to downsample the output of the convolutional layers. Pooling layers are typically applied after each convolutional layer, and they reduce the spatial resolution of the feature maps produced by the convolutional layer. This has the effect of increasing the receptive field of the filters, which allows the network to learn features that are invariant to small translations and deformations in the input data.

\paragraph{Fully Connected Layers}
After the convolutional and pooling layers, the output of the CNN is passed through a series of fully connected layers, which combine the features extracted by the convolutional and pooling layers to produce a final output. The weights of the filters and fully connected layers are learned from the data during training, which allows the network to learn the patterns and structures present in the data. Once the network has been trained, it can be applied to new data to produce a segmentation map.

\par In the context of medical image segmentation, CNNs are able to learn the complex patterns and structures present in medical images, which allows them to accurately segment different structures and objects within the images. For example, a CNN can learn to identify different organs, tissues, and lesions in medical images, which can help doctors diagnose and treat medical conditions.

\subsection{Underfitting/Overfitting}\label{overfittingunderfitting}
Underfitting and overfitting are common challenges in CNNs for medical image segmentation. Underfitting occurs when a CNN is not complex enough to capture the patterns and structures present in the data, which can lead to poor performance on the task. On the other hand, overfitting occurs when a CNN is too complex relative to the amount of data available, which can cause the network to learn patterns that are specific to the training data but do not generalize to new data. Both underfitting and overfitting can negatively impact the performance of a CNN for medical image segmentation.

\subsubsection{Diagnosing}
One way to diagnose underfitting and overfitting in a CNN for medical image segmentation is by analyzing the performance of the network on the training set and the validation set. The training set is the data used to train the network, while the validation set is a separate set of data that is held out from training and used to evaluate the performance of the network. If the performance of the network on the training set is significantly better than the performance on the validation set, this may indicate overfitting, as the network is able to learn patterns that are specific to the training data but do not generalize to new data. On the other hand, if the performance of the network is poor on both the training set and the validation set, this may indicate underfitting, as the network is not complex enough to capture the patterns and structures present in the data.
\begin{figure}[h]
    \centering \includegraphics[width=0.5\textwidth]{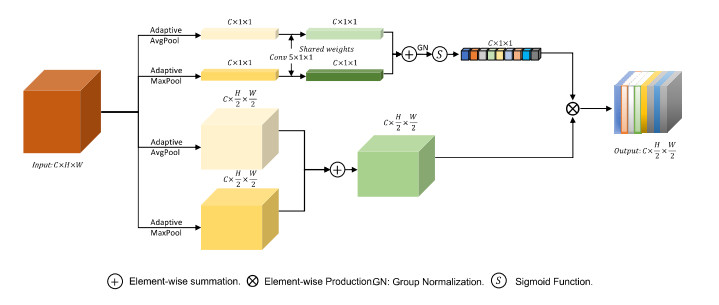}
    \caption{+ Element-wise summation x-Element-wise production GN: Group-wise Normalisation \textit{S}:Sigmoid function}
\end{figure}
 
\subsection{Trainable Parameters}
The number of trainable parameters in a CNN plays an important role in image segmentation, as it determines the capacity of the network to learn from the data. In general, a larger number of trainable parameters allows the network to learn more complex patterns and structures from the data, which can improve the accuracy of the segmentation map. However, it is also important to ensure that the network has enough data to learn from, as a large number of trainable parameters without sufficient data can result in overfitting, which can reduce the generalizability of the network.

In the context of image segmentation, the number of trainable parameters in a CNN can determine the spatial resolution of the segmentation map that the network is able to produce. For example, a CNN with a large number of trainable parameters may be able to produce a segmentation map with fine-grained details, while a CNN with a smaller number of trainable parameters may produce a coarser segmentation map. Therefore, it is important to carefully balance the number of trainable parameters in a CNN with the amount of data available for training in order to achieve the best performance on the task at hand.

One way to control the number of trainable parameters in a CNN is by adjusting the size of the filters used in the convolutional layers. Larger filters have more trainable parameters, which allows the network to learn more complex patterns from the data, but they also require more data to learn from in order to avoid overfitting. On the other hand, smaller filters have fewer trainable parameters, which reduces the risk of overfitting, but they may not be able to capture as many fine-grained details in the data.

Overall, the number of trainable parameters in a CNN is an important factor to consider when designing a network for image segmentation. By carefully balancing the number of trainable parameters with the amount of data available, it is possible to achieve good performance on the task while avoiding overfitting. Advantages of reducing trainable
parameters:\label{optimizingparameters}

\begin{itemize}
    \item First, reducing the number of trainable parameters can reduce the risk of overfitting, which can improve the generalizability of the network and allow it to perform well on unseen data. By reducing the number of trainable parameters, it is possible to reduce the risk of overfitting and improve the performance of the network on unseen data.
    \item Second, reducing the number of trainable parameters can also make the network more efficient, both in terms of training time and inference time. As the number of trainable parameters increases, the network becomes more complex and requires more data to learn from, which can increase the time and computational resources required to train the network. By reducing the number of trainable parameters, it is possible to reduce the complexity of the network and make it more efficient to train and deploy.
    \item Third, reducing the number of trainable parameters can also make the network more interpretable, which can be important in medical applications. As the number of trainable parameters increases, the network becomes more complex and difficult to understand, which can make it challenging to interpret the results of the network. By reducing the number of trainable parameters, it is possible to make the network more interpretable and better understand how the network is making its predictions.
\end{itemize}

\section{Literature Review}
\noindent Medical image segmentation must be precise for processing \& analysis to be successful. In the past, this was often done manually by doctors, but this was time-consuming and required specialized knowledge. Many automatic segmentation methods have been proposed, based on techniques such as edge detection \cite{IJSRSET} and template matching \cite{8477639}, but these can be difficult to adapt to different segmentation tasks. The variability in the size and shape of the targets also presents challenges for segmentation.
{Approaching image segmentation with CNNs}
There are many different research papers and articles that have explored the use of CNNs for medical image segmentation. For example, one study used a CNN to segment liver tumors from CT scans \cite{app12178650}, and achieved promising results compared to traditional image processing methods. Another study \cite{Xu2022-sv} used a CNN to segment the left ventricle of the heart from Cardiac MR images, and showed that it was able to produce more accurate and consistent results than previous methods.

\subsubsection{U-Net}\label{unetsection}
\begin{figure}[h]
    \centering \includegraphics[width=0.5\textwidth]{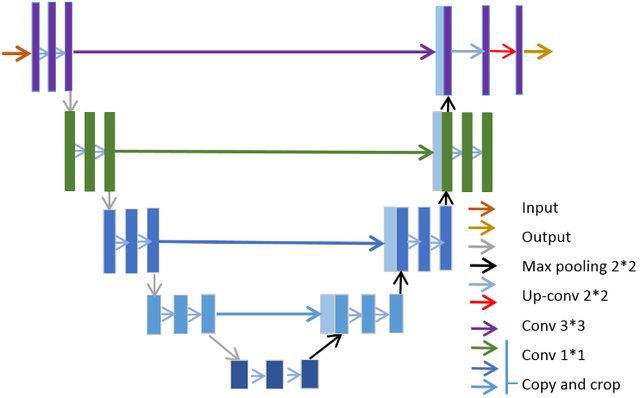}
    \caption{U-Net architecture. Ref: \cite{unet}}
\end{figure}

U-Net \cite{unet} is a CNN that is composed of a contracting path and an expansive path. The contracting path is a traditional CNN that is used to extract features from the input image, while the expansive path is used to upsample the output of the contracting path to generate a segmentation map.

\par The architecture of U-Net is based on the encsuboder-decoder structure, where the contracting path acts as the encoder and the expansive path acts as the decoder. The encoder is composed of a series of convolutional layers, each of which reduces the spatial resolution of the input image by a factor of two. The decoder is composed of a series of transposed convolutional layers, each of which increases the spatial resolution of the output by a factor of two.

\par One of the main contribution of U-Net is its use of skip connections between the encoder and decoder to improve the accuracy of the segmentation map. These skip connections allow the network to combine high-level features from the encoder with low-level features from the decoder, which helps improve the performance of the network. Prior to the development of U-Net, many image segmentation algorithms relied on post-processing techniques to improve the accuracy of their segmentation maps as mentioned in section \ref{thresholdingmethod} and \ref{clusteringmethod}, but U-Net was the first architecture to incorporate skip connections as a means of improving segmentation performance.

\par U-Net is known for its performance in medical image segmentation tasks, but it can be used for other types of image segmentation as well. It has been applied to a wide range of tasks, including cell segmentation \cite{cellsegmentation}, brain tumor segmentation \cite{braintumorsegmentation}, and satellite image segmentation \cite{rs13040808}.

\subsubsection{Attention mechanism in U-Net}
The attention mechanism in U-Net is a way of weighting the different features extracted by the convolutional layers of the network, based on their importance for the segmentation task. This is done using a self-attention mechanism, where each feature is multiplied by a scalar weight that reflects its contribution to the final segmentation \cite{2018arXiv180403999O}.

The attention mechanism in U-Net is implemented as a series of additional convolutional layers, called attention gates, that are inserted between the encoder and decoder portions of the network. These layers use the features extracted by the encoder to compute the attention weights, and then apply these weights to the corresponding features in the decoder. This allows the network to focus on the most relevant features for the segmentation task, and to suppress the effects of noise and other irrelevant features. It can help the network to make more accurate and consistent predictions, and to handle images with complex intensity distributions or multiple classes of pixels.

In NLP tasks for machine translation, \cite{nlpattention} utilised attention mechanisms. CA-Net \cite{9246575} developed a thorough attention network using multi-scale feature maps to highlight significant features. For more detailed context information, Attention U-Net \cite{2018arXiv180403999O} used attention gates. These techniques contain more parameters but benefit from context information to increase accuracy.

\subsubsection{Residual Connections in U-Net}
The research MultiResU-Net \cite{IBTEHAZ202074} uses multiple resolution scales to process the input data. In other words, the input image is processed at multiple scales, with each scale being processed by a separate U-Net sub-network. The output of each sub-network is then combined to produce the final segmentation result. This allows the network to capture both fine-grained and coarse-grained information from the input image, which can improve the accuracy of the segmentation.

MultiResU-Net has been shown to outperform other state-of-the-art methods at the time (2020) for medical image segmentation on a range of tasks, including segmenting cells in microscopy images and segmenting organs in CT and MRI scans.

\paragraph{Problem of vanishing gradients}
The vanishing gradient problem is a common issue in deep learning, where the gradients of the parameters with respect to the loss function become very small during training. This can happen in medical image segmentation, where the network is trying to learn complex patterns in the images to perform the segmentation.

When the gradients become very small, the network has a difficulty in learning because the weights of the network are updated using the gradients. If the gradients are very small, the weights are not updated effectively, and thus the difficulty in learning the patterns in the data.

To address the vanishing gradient problem, researchers have developed various techniques, such as using deeper networks as mentioned in section \ref{resnetsection}, using skip connections as mention in section \ref{unetsection}, and using different activation functions \cite{WANG201988}.

\paragraph{Residual Network}\label{resnetsection}

\begin{figure}[h]
    \centering \includegraphics[width=0.5\textwidth]{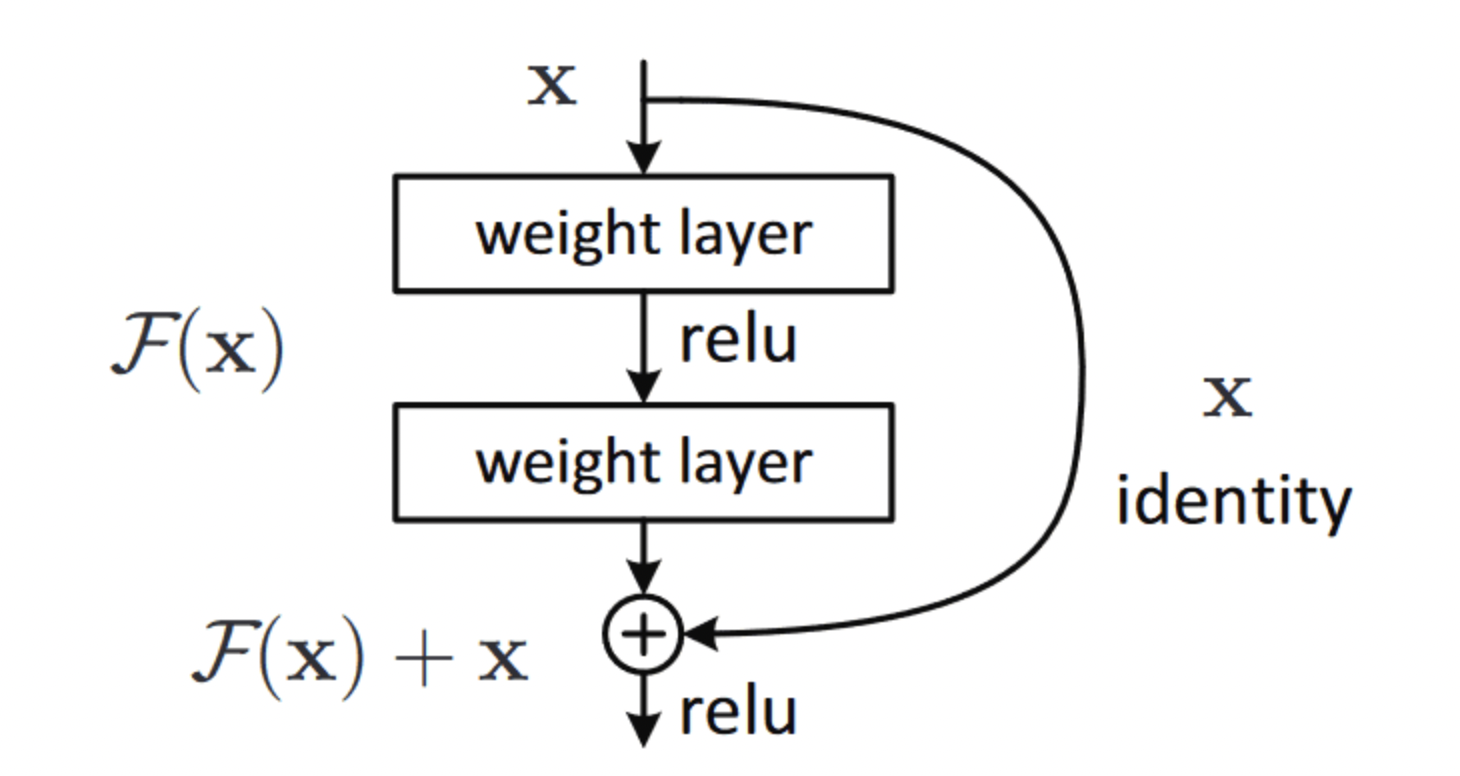}
    \caption{Residual learning. Ref: \cite{resnetpaperssss}}
\end{figure}

Residual Network (ResNet) was designed to address the problem of vanishing gradients. In a traditional neural network, each layer receives input from the previous layer, applies a series of transformations to the input, and then passes the transformed data on to the next layer. However, as the number of layers in a neural network increases, it becomes increasingly difficult for the network to train effectively. This is because the signal that is passed from one layer to the next can become very weak, making it difficult for the network to learn from the data.

ResNet addresses this problem by introducing a new type of layer called a residual layer. In a residual layer, instead of just passing the transformed data to the next layer, the residual layer also adds the original input data to the transformed data before passing it on. This allows the network to learn much deeper networks, because it enables the signal to be preserved even as it passes through many layers.

The use of residual layers has been shown to greatly improve the performance of neural networks on a wide range of tasks, and ResNet has become a widely used architecture for deep learning. It has been used in applications such as image classification \cite{SARWINDA2021423}, object detection \cite{resnetobjectdetection}, et cetera.

\section{Proposed methodology}
\subsection{Depthwise Separable Convolutions}
\begin{figure}[h]
    \centering \includegraphics[width=0.5\textwidth]{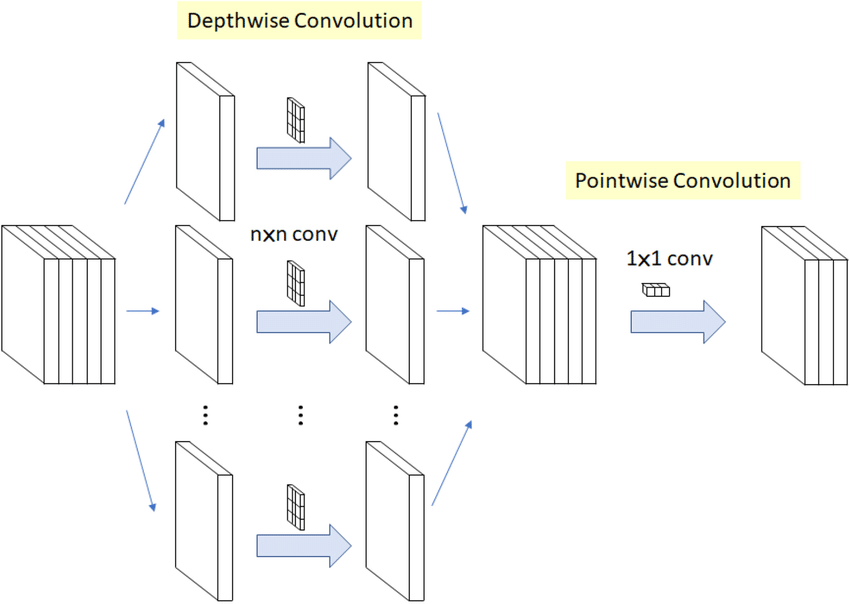}
    \caption{Depthwise + Pointwise convolution. Ref: \cite{depthwiseconv}}
\end{figure}

Depthwise separable convolution is a type of convolutional operation that reduces the number of parameters and computations in the neural network.

In medical image segmentation, depthwise separable convolutions can be used to efficiently process 3D medical images, such as MRI or CT scans. These images typically have a large number of channels, corresponding to the different tissue types or other features in the body. The algorithm is composed of two steps:
\subsubsection{Depthwise Convolution}
In the depthwise convolution step, a single convolution filter is applied to each input channel. This filter operates upon the entire feature map, so it has a filter size equal to the feature map size. This step is used to separate the feature maps into distinct channels.

\subsubsection{Pointwise Convolution}
In the pointwise convolution step, a 1x1 convolution filter is applied to each of the input channels. This filter is used to combine the feature maps from the depthwise convolution step into a single output feature map. It is also used to reduce the number of input channels to the desired number of output channels.

\subsection{Attention in Skip connections}

\subsubsection{Hard Attention}
Hard attention is a type of attention mechanism that uses a binary mask to focus the model's attention on specific parts of the image. The mask is generated by a separate network, which receives the input image and outputs the binary mask. The mask is then applied to the input image, zeroing out all pixels that are not part of the focus area. The model then processes the masked image, using only the remaining pixels to make predictions.

\subsubsection{Soft Attention}
\begin{figure}[h]
    \centering \includegraphics[width=0.5\textwidth]{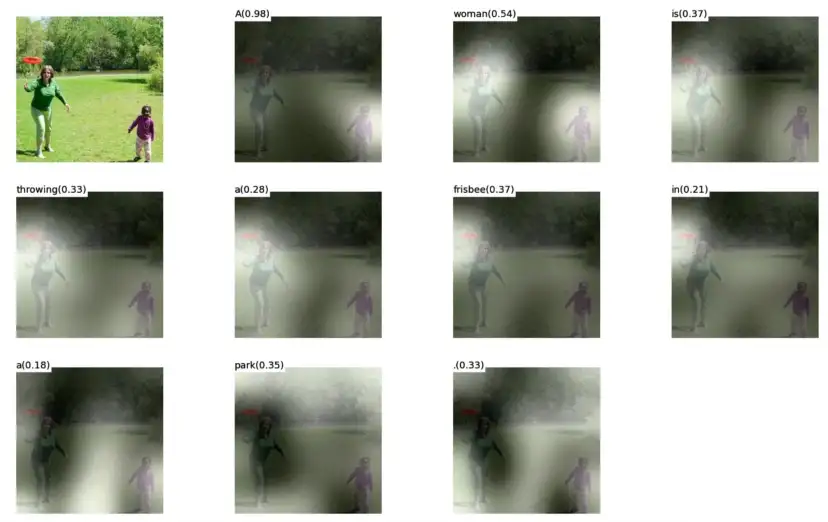}
    \caption{Soft attention weights visualisation. Ref: \cite{2018arXiv180403999O}}
\end{figure}

Soft attention is a type of attention mechanism that uses a continuous weighting function to focus the model's attention on specific parts of the image. The weighting function is generated by a separate network, which receives the input image and outputs a set of weights. These weights are then applied to the input image, effectively giving each pixel a different degree of importance. The model then processes the weighted image, taking into account the different importance of each pixel when making predictions.

\subsubsection{Comparison}
Overall, hard attention is a more coarse-grained approach to focusing the model's attention, as it only allows the model to attend to a binary (on or off) set of pixels. 

Soft attention, on the other hand, is a more fine-grained approach, as it allows the model to attend to a continuous range of pixel intensities.

\subsubsection{Implementation}
\begin{figure}[h]
    \centering \includegraphics[width=0.5\textwidth]{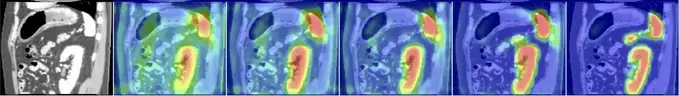}
    \caption{A higher attention coefficient is highlighted in red when it is visualised at 3, 6, 10, and 150 epochs. Ref: \cite{2018arXiv180403999O}}
\end{figure}

To implement soft attention in the skip connections of a U-Net, a separate network is used to generate the weighting function for each skip connection, as explained in \cite{2018arXiv180403999O}. This weighting function is then applied to the features coming from the lower-level part of the network, effectively giving each feature a different degree of importance. The weighted features are then combined with the features coming from the higher-level part of the network, and the resulting combined features are passed through the next layer of the network. This process is repeated for each skip connection in the U-Net, allowing the model to automatically learn which features to combine and how to combine them in order to make more accurate predictions.

\subsection{Overall Architecture}
\begin{figure}[h]
    \centering \includegraphics[width=0.5\textwidth]{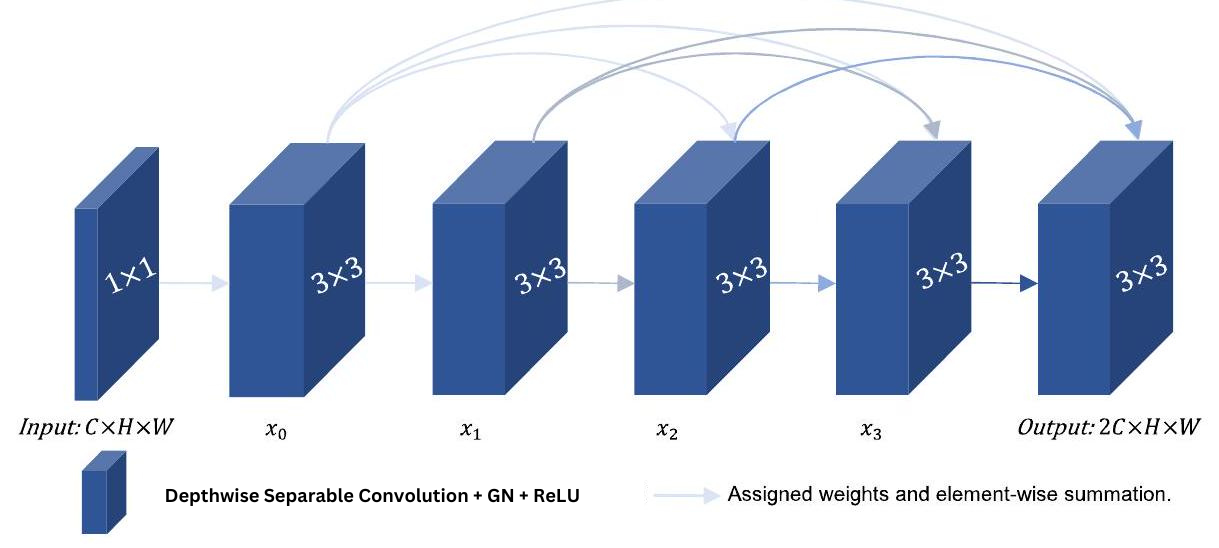}
    \caption{Basic CNN block of proposed architecture}
\end{figure}
The proposed U-Net architecture incorporates attention mechanism and densely connected blocks. The residual connections are used in this to help with the flow of gradients during backpropagation and improve the learning of deeper features. Here is the overall architecture:

\begin{itemize}
    \item Encoder: The encoder part of the network consists of multiple densely connected blocks, where each block contains a series of convolutional layers followed by batch normalization and ReLU activation. The output of each block is passed through a residual connection to the next block, which helps with gradient flow and enables the network to learn deeper features.
    \item Attention gate: At each level of the encoder, an attention gate is added that takes as input the feature maps of the corresponding encoder block and the feature maps of the decoder block. The attention gate computes a set of attention coefficients that highlight the relevant features in the encoder maps for the decoder.
    \item Decoder: The decoder part of the network is similar to the encoder, but instead of dense blocks, it uses upsampling layers and skip connections to combine feature maps from the encoder with the upsampled feature maps of the decoder. The feature maps are passed through another set of densely connected blocks before being fed to the next level of the decoder.
    \item Output: The output of the decoder is a segmentation map that corresponds to the input image. The final layer of the decoder is a 1x1 convolutional layer that maps the feature maps to the number of classes in the segmentation task.
\end{itemize}

\subsubsection{Basic CNN block}
Figure 5.5 illustrates that each depthwise separable convolution layer is followed by group normalization and LeakyReLU to enhance the model's nonlinear expression capability. Additionally, residual connections from previous layers are combined with subsequent layers through elementwise summation, which uses no extra parameters to fuse extracted information. It is described by the following equation:

\begin{equation}
x_{0}=F_{1 \times 1}^{\operatorname{conv}}\left(x_{\text {input }}\right) \\
\end{equation}
\begin{equation}
x_{i}=F_{3 \times 3}^{\text {conv }}\left(\operatorname{SUM}\left(x_{0} ; x_{1} ; x_{2} ; \ldots ; x_{i-1}\right)\right)
\end{equation}

where $x_{i} \in \mathbb{R}^{C \times H \times W}$ represents feature maps in layer $i$, and the feature map is represented with $x_{0} \in \mathbb{R}^{C \times H \times W}$.

\subsubsection{Pooling Layer with Attention}
The attention mechanism can be implemented in various ways, but a common approach is to use a CNN to learn an attention map for the input image \cite{9156697}. The attention map is a 2D grid of values, where each value indicates the importance of a corresponding region of the input image.

The pooling operation is then performed by taking the weighted sum of the input image, where the weights are determined by the attention map. This allows the model to focus on the most important parts of the input image when performing the pooling, and can help to improve the accuracy of the model.

\subsection{Dataset}
\subsection{Skin Lesion Segmentation}
The HAM10000 dataset \cite{Tschandl2018} is a massive collection of dermoscopic images of pigmented skin lesions from various sources. Like many other medical imagery datasets, there is class imbalance in this dataset, with only 115 images of a certain class among 10015 total images of size 450 x 600.

\subsubsection{Categories of skin lesions}
\begin{figure}[h]
    \centering \includegraphics[width=0.5\textwidth]{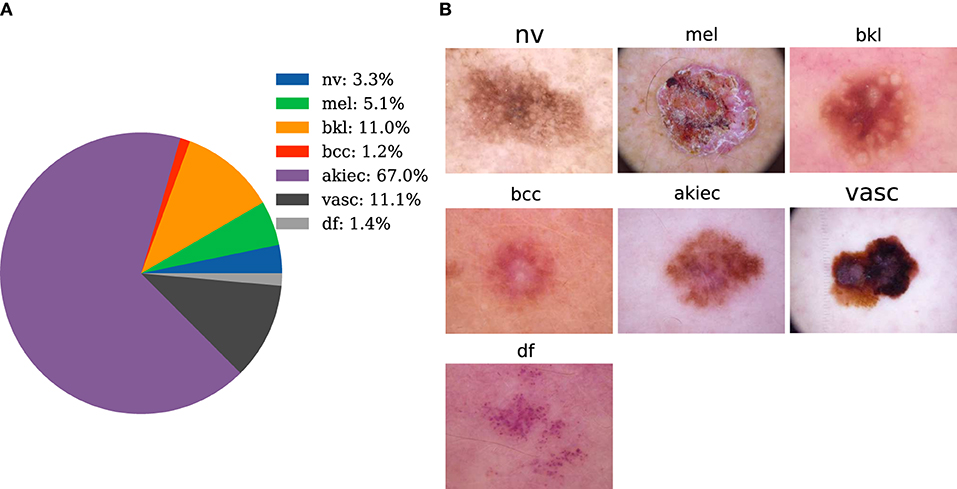}
    \caption{HAM10000 dataset distribution. Ref: \cite{Tschandl2018}}
\end{figure}

\noindent There are seven types of skin lesions:
\begin{enumerate}
    \item Actinic Keratosis/Intraepithelial Carcinoma (AKIEC) - 327 images
    \item Basal Cell Carcinoma (BCC) - 514 images
    \item Benign Keratosis (BKL) - 1099 images
    \item Dermatofibroma (DF) - 115 images
    \item Melanoma (MEL) - 1113 images
    \item Melanocytic Nevi (NV) - 6705 images
    \item Vascular lesions (VASC) - 142 images
\end{enumerate}

\subsubsection{Pre-processing}
All images are cropped to a standard size of 256 x 256. The training, testing and validation sets are segregated:
\begin{itemize}
    \item Training: 80\% (~8k images)
    \item Testing: 10\% (~1k images)
    \item Validation: 10\% (~1k images)
\end{itemize}


\subsection{Metrics}\label{metrics}
\subsubsection{Dice coefficient}
The Dice Coefficient is a measure of similarity between two binary images. It is used to quantify the amount of overlap between two images and is used to assess the accuracy of medical image segmentation. Mathematically, it is expressed as follows:
\begin{center}
$Dice = \frac{2*TP}{2TP+FN+FP}$\\
\end{center}

The dice coefficient is a value between 0 and 1, where a value of 1 indicates perfect overlap between the predicted and ground truth segmentations.

\subsubsection{Intersection over Union (IOU)}
The IOU metric is used to evaluate the performance of a medical image segmentation algorithm. This is calculated by dividing the area of overlap between the ground truth segmentation and the predicted segmentation by the area of union of both segmentations. It is represented mathematically as follows:
\begin{center}
$IoU = \frac{TP}{TP+FN+FP}$\\
\end{center}

The IoU ranges from 0 to 1, where 0 means no overlap and 1 means perfect overlap. A higher IoU indicates a better accuracy of the segmentation algorithm.

\subsubsection{Average Symmetric Surface Distance}
Average symmetric surface distance (ASSD) is a metric used to measure the average distance between the surfaces of two objects in a 3D image, such as the surface of a tumor or organ and the surrounding tissue. It is mathematically defined as follows:
\begin{center}
$ASSD = \frac{1}{N}\sum{d(p, q)}$    
\end{center}

where N is the number of points on the predicted segmentation boundary, d(p,q) is the distance between a point p on the predicted boundary and its nearest point q on the ground truth boundary, and the sum is taken over all N points on the predicted boundary.

The smaller the ASSD, the more accurate the predicted segmentation is.

\section{Results}
In order to evaluate the proposed network, the architecture was validated and compared against other state-of-the-art methods using two publicly available datasets: one focused on skin lesion segmentation \cite{Tschandl2018} and the other on thyroid gland segmentation \cite{thyroidsegmentationpaper}. The outcomes achieved by other top-performing networks were compared for each task, and ablation studies were conducted to demonstrate the effectiveness of the various modules used.

\subsection{Skin Lesion Segmentation}\label{skinlesionsection}

\begin{figure}[h]
    \centering \includegraphics[width=0.4\textwidth]{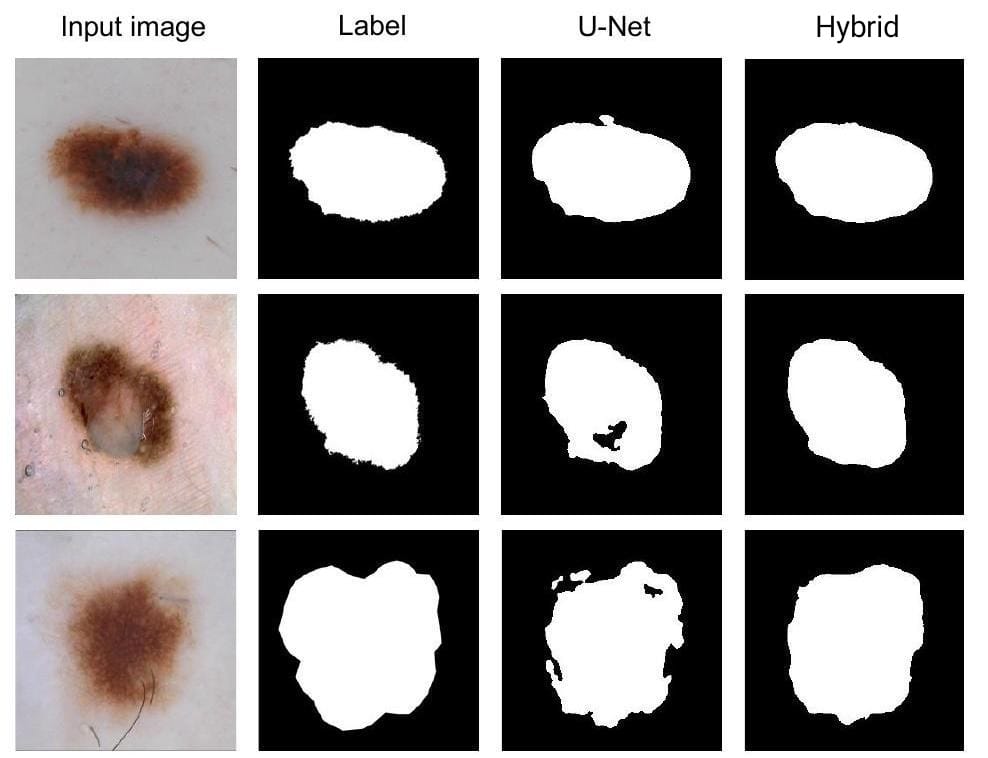}
    \caption{Results of visualization for skin lesion segmentation}
\end{figure}

\begin{table}[h]
\centering
\caption{Ablation study for HAM1000 \cite{Tschandl2018} dataset}
\begin{tabular}{llll}
\hline
Components & IoU & Dice & ASSD \\
\hline
U-Net + Depthwise Convolution (DC) & 0.7164 & 0.8232 & 1.6189 \\
U-Net + DC + Residual Connections (RC) & 0.7672 & 0.8612 & 1.0833 \\
U-Net + DC + RC + Attention Pooling & 0.8157 & 0.8872 & 0.8364 \\
\hline
\end{tabular}
\end{table}

Figure 6.1 presents skin lesion segmentation samples for the proposed hybrid network and U-Net. The proposed approach outperforms U-Net in regular skin lesion segmentation images, particularly when the skin lesion has similar color to its surroundings or is occluded by hair and tissue fluid. U-Net produces incorrect segmentation results when the boundary of the skin lesion is more blurred. To validate the method, U-Net \cite{unet}, Attention U-Net \cite{2018arXiv180403999O}, and MultiResUNet \cite{IBTEHAZ202074} are compared, and the results are presented in Table 6.1. The proposed model achieved higher accuracy, dice coefficient, and lower number of trainable parameters than U-Net and Attention U-Net. Although MultiResUNet has better metrics, the hybrid approach reduces the number of parameters by almost 97\%.

\begin{table}[h]
\centering
\caption{Comparison of hybrid approach to other approaches for skin lesion segmentation}
\begin{tabular}{llll}
\hline
Approach & Accuracy & Dice Coefficient & Number of Params \\
\hline
U-Net \cite{unet} & 0.8777 & 0.8739 & $7.76 \mathrm{M}$ \\
Attention-UNet \cite{2018arXiv180403999O} & 0.8812 & 0.8854 & $34.88 \mathrm{M}$ \\
MultiResUNet \cite{IBTEHAZ202074} & \textbf{0.9221} & \textbf{0.9179} & $64.8 \mathrm{ M}$ \\
\textbf{Hybrid (ours)} & 0.9082 & 0.8872 & \textbf{2.3}$\mathrm{\textbf{M}}$ \\
\hline
\end{tabular}
\end{table}

\subsection{Thyroid Gland Segmentation}
\begin{figure}[h]
    \centering \includegraphics[width=0.5\textwidth]{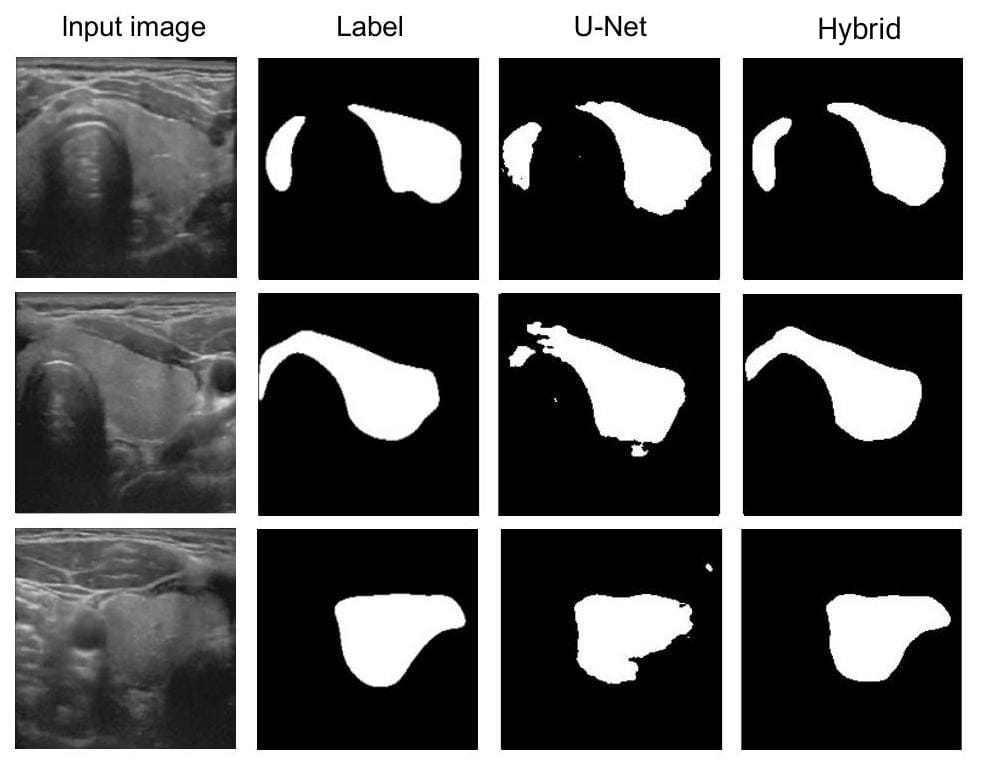}
    \caption{Results of visualization for thyroid gland segmentation}
\end{figure}

Several results of segmentations on the thyroid gland dataset \cite{thyroidsegmentationpaper} are displayed in Figure 6.2. U-Net produces an insufficient segmentation of the thyroid isthmus, while the hybrid approach produced better results. Same base approaches are used to compare results on the thyroid gland dataset, as mentioned in skin lesion segmentation 
 \ref{skinlesionsection}.

\begin{table}[h]
\centering
\caption{Comparison of hybrid approach to other approaches for thyroid gland segmentation}
\begin{tabular}{llllc}
\hline
Approach & Accuracy & Dice Coefficient & Number of Params \\
\hline
U-Net \cite{unet} & 0.9347 & 0.9332 & $7.76 \mathrm{M}$ \\
Attention-UNet \cite{2018arXiv180403999O} & 0.9526 & 0.9169 & $34.88 \mathrm{M}$ \\
MultiResUNet \cite{IBTEHAZ202074} & 0.9727 & \textbf{0.9544} & $64.8 \mathrm{M}$ \\
\textbf{Hybrid (ours)} & \textbf{0.9766} & 0.9493 & \textbf{2.3}$\mathrm{\textbf{M}}$ \\
\hline
\end{tabular}
\end{table}

\begin{table}[!h]
\centering
\caption{Ablation study for thyroid gland dataset \cite{thyroidsegmentationpaper}}
\begin{tabular}{llll}
\hline
Components & IoU & Dice & ASSD \\
\hline
U-Net + Depthwise Convolution (DC) & 0.8079 & 0.8837 & 1.0703 \\
U-Net + DC + Residual Connections (RC) & 0.8325 & 0.9017 & 0.8331 \\
U-Net + DC + RC + Attention Pooling & 0.9544 & 0.9493 & 0.0953 \\
\hline
\end{tabular}
\end{table}

\section{Conclusion}
There are many different techniques and algorithms used for medical image segmentation, including thresholding, clustering, and deep learning methods. In this research, we improve approaches like MultiResUNet \cite{IBTEHAZ202074}, Attention U-Net \cite{2018arXiv180403999O}, classical U-Net \cite{unet}, and other variants, to ultimately perform better on some medical image segmentation tasks such as skin lesion segmentation \cite{Tschandl2018} and thyroid gland segmentation \cite{thyroidsegmentationpaper}. To assess the effectiveness of the proposed network, its architecture was verified and contrasted with other advanced techniques using two publicly accessible datasets: one dataset primarily concerned with identifying skin lesions and the other dataset focused on identifying the thyroid gland.

\begin{itemize}
    \item Depthwise separable convolutions are used in the encoder to reduce the number of parameters and improve the efficiency of the network
    \item Attention mechanism is used to selectively emphasize important features in encoder maps for the decoder, which helps to improve the accuracy of segmentation
    \item Residual connections are used to help with the flow of gradients during backpropagation
\end{itemize}

Overall, the combination of these techniques results in a more accurate and efficient segmentation network, with state-of-the-art performance on various segmentation tasks.




\bibliographystyle{IEEEtran}
\bibliography{Hybrid_Approach}

\begin{thebibliography}{10}
\providecommand{\url}[1]{#1}
\csname url@samestyle\endcsname
\providecommand{\newblock}{\relax}
\providecommand{\bibinfo}[2]{#2}
\providecommand{\BIBentrySTDinterwordspacing}{\spaceskip=0pt\relax}
\providecommand{\BIBentryALTinterwordstretchfactor}{4}
\providecommand{\BIBentryALTinterwordspacing}{\spaceskip=\fontdimen2\font plus
\BIBentryALTinterwordstretchfactor\fontdimen3\font minus
  \fontdimen4\font\relax}
\providecommand{\BIBforeignlanguage}[2]{{%
\expandafter\ifx\csname l@#1\endcsname\relax
\typeout{** WARNING: IEEEtran.bst: No hyphenation pattern has been}%
\typeout{** loaded for the language `#1'. Using the pattern for}%
\typeout{** the default language instead.}%
\else
\language=\csname l@#1\endcsname
\fi
#2}}
\providecommand{\BIBdecl}{\relax}
\BIBdecl

\bibitem{9246575}
R.~Gu, G.~Wang, T.~Song, R.~Huang, M.~Aertsen, J.~Deprest, S.~Ourselin,
  T.~Vercauteren, and S.~Zhang, ``Ca-net: Comprehensive attention convolutional
  neural networks for explainable medical image segmentation,'' \emph{IEEE
  Transactions on Medical Imaging}, vol.~40, no.~2, pp. 699--711, 2021.

\bibitem{IBTEHAZ202074}
N.~Ibtehaz and M.~S. Rahman, ``Multiresunet : Rethinking the u-net architecture
  for multimodal biomedical image segmentation,'' \emph{Neural Networks}, vol.
  121, pp. 74--87, 2020.

\bibitem{Tschandl2018}
P.~Tschandl, C.~Rosendahl, and H.~Kittler, ``The ham10000 dataset, a large
  collection of multi-source dermatoscopic images of common pigmented skin
  lesions,'' \emph{Scientific Data}, vol.~5, no.~1, p. 180161, Aug 2018.

\bibitem{xceptionpaper}
F.~Chollet, ``Xception: Deep learning with depthwise separable convolutions,''
  07 2017, pp. 1800--1807.

\bibitem{2018arXiv180403999O}
O.~{Oktay}, J.~{Schlemper}, L.~{Le Folgoc}, M.~{Lee}, M.~{Heinrich},
  K.~{Misawa}, K.~{Mori}, S.~{McDonagh}, N.~{Y Hammerla}, B.~{Kainz},
  B.~{Glocker}, and D.~{Rueckert}, ``{Attention U-Net: Learning Where to Look
  for the Pancreas},'' \emph{arXiv e-prints}, p. arXiv:1804.03999, Apr. 2018.

\bibitem{GOH2018298}
T.~Y. Goh, S.~N. Basah, H.~Yazid, M.~J. {Aziz Safar}, and F.~S. {Ahmad Saad},
  ``Performance analysis of image thresholding: Otsu technique,''
  \emph{Measurement}, vol. 114, pp. 298--307, 2018.

\bibitem{9356353}
S.~Minaee, Y.~Boykov, F.~Porikli, A.~Plaza, N.~Kehtarnavaz, and D.~Terzopoulos,
  ``Image segmentation using deep learning: A survey,'' \emph{IEEE Transactions
  on Pattern Analysis and Machine Intelligence}, vol.~44, no.~7, pp.
  3523--3542, 2022.

\bibitem{FU2021107}
Y.~Fu, Y.~Lei, T.~Wang, W.~J. Curran, T.~Liu, and X.~Yang, ``A review of deep
  learning based methods for medical image multi-organ segmentation,''
  \emph{Physica Medica}, vol.~85, pp. 107--122, 2021.

\bibitem{braintumorsegmentation}
G.~Wang, W.~Li, S.~Ourselin, and T.~Vercauteren, ``Automatic brain tumor
  segmentation using convolutional neural networks with test-time
  augmentation,'' in \emph{Brainlesion: Glioma, Multiple Sclerosis, Stroke and
  Traumatic Brain Injuries}, A.~Crimi, S.~Bakas, H.~Kuijf, F.~Keyvan, M.~Reyes,
  and T.~van Walsum, Eds.\hskip 1em plus 0.5em minus 0.4em\relax Cham: Springer
  International Publishing, 2019, pp. 61--72.

\bibitem{IJSRSET}
{Rana Riad K. Al-Taie}, { Basma Jumaa Saleh}, and { Lamees Abdalhasan Salman},
  ``Image edge-segmentation techniques : A review,'' \emph{International
  Journal of Scientific Research in Science, Engineering and Technology},
  vol.~8.

\bibitem{8477639}
K.~Yogheedha, A.~Nasir, H.~Jaafar, and S.~Mamduh, ``Automatic vehicle license
  plate recognition system based on image processing and template matching
  approach,'' in \emph{2018 International Conference on Computational Approach
  in Smart Systems Design and Applications (ICASSDA)}, 2018, pp. 1--8.

\bibitem{app12178650}
M.~W. Sabir, Z.~Khan, N.~M. Saad, D.~M. Khan, M.~A. Al-Khasawneh, K.~Perveen,
  A.~Qayyum, and S.~S. Azhar~Ali, ``Segmentation of liver tumor in ct scan
  using resu-net,'' \emph{Applied Sciences}, vol.~12, no.~17, 2022.

\bibitem{Xu2022-sv}
S.~Xu, H.~Lu, S.~Cheng, and C.~Pei, ``\BIBforeignlanguage{en}{Left ventricle
  segmentation in cardiac {MR} images via an improved {ResUnet}},''
  \emph{\BIBforeignlanguage{en}{Int J Biomed Imaging}}, vol. 2022, p. 8669305,
  Jul. 2022.

\bibitem{unet}
O.~Ronneberger, P.~Fischer, and T.~Brox, ``U-net: Convolutional networks for
  biomedical image segmentation,'' vol. 9351, 10 2015, pp. 234--241.

\bibitem{cellsegmentation}
N.~Greenwald, G.~Miller, E.~Moen, A.~Kong, A.~Kagel, T.~Dougherty, C.~Fullaway,
  B.~McIntosh, K.~Leow, M.~Schwartz, C.~Pavelchek, S.~Cui, I.~Camplisson,
  O.~Bar-Tal, J.~Singh, M.~Fong, G.~Chaudhry, Z.~Abraham, J.~Moseley, and
  D.~Valen, ``Whole-cell segmentation of tissue images with human-level
  performance using large-scale data annotation and deep learning,''
  \emph{Nature Biotechnology}, vol.~40, 04 2022.

\bibitem{rs13040808}
B.~Neupane, T.~Horanont, and J.~Aryal, ``Deep learning-based semantic
  segmentation of urban features in satellite images: A review and
  meta-analysis,'' \emph{Remote Sensing}, vol.~13, no.~4, 2021.

\bibitem{nlpattention}
D.~Bahdanau, K.~Cho, and Y.~Bengio, ``Neural machine translation by jointly
  learning to align and translate,'' \emph{ArXiv}, vol. 1409, 09 2014.

\bibitem{WANG201988}
X.~Wang, Y.~Qin, Y.~Wang, S.~Xiang, and H.~Chen, ``Reltanh: An activation
  function with vanishing gradient resistance for sae-based dnns and its
  application to rotating machinery fault diagnosis,'' \emph{Neurocomputing},
  vol. 363, pp. 88--98, 2019.

\bibitem{resnetpaperssss}
K.~He, X.~Zhang, S.~Ren, and J.~Sun, ``Deep residual learning for image
  recognition,'' 06 2016, pp. 770--778.

\bibitem{SARWINDA2021423}
D.~Sarwinda, R.~H. Paradisa, A.~Bustamam, and P.~Anggia, ``Deep learning in
  image classification using residual network (resnet) variants for detection
  of colorectal cancer,'' \emph{Procedia Computer Science}, vol. 179, pp.
  423--431, 2021, 5th International Conference on Computer Science and
  Computational Intelligence 2020.

\bibitem{resnetobjectdetection}
Z.~Bai and D.~Jiang, ``On the multi-scale real-time object detection using
  resnet,'' in \emph{Pattern Recognition and Computer Vision}, Z.~Lin, L.~Wang,
  J.~Yang, G.~Shi, T.~Tan, N.~Zheng, X.~Chen, and Y.~Zhang, Eds.\hskip 1em plus
  0.5em minus 0.4em\relax Cham: Springer International Publishing, 2019, pp.
  63--73.

\bibitem{depthwiseconv}
I.~Junejo and N.~Ahmed, ``Depthwise separable convolutional neural networks for
  pedestrian attribute recognition,'' \emph{SN Computer Science}, vol.~2, 04
  2021.

\bibitem{9156697}
Q.~Wang, B.~Wu, P.~Zhu, P.~Li, W.~Zuo, and Q.~Hu, ``Eca-net: Efficient channel
  attention for deep convolutional neural networks,'' in \emph{2020 IEEE/CVF
  Conference on Computer Vision and Pattern Recognition (CVPR)}, 2020, pp.
  11\,531--11\,539.

\bibitem{thyroidsegmentationpaper}
\BIBentryALTinterwordspacing
T.~Wunderling, B.~Golla, P.~Poudel, C.~Arens, M.~Friebe, and C.~Hansen,
  ``{Comparison of thyroid segmentation techniques for 3D ultrasound},'' in
  \emph{Medical Imaging 2017: Image Processing}, M.~A. Styner and E.~D.
  Angelini, Eds., vol. 10133, International Society for Optics and
  Photonics.\hskip 1em plus 0.5em minus 0.4em\relax SPIE, 2017, p. 1013317.
  [Online]. Available: \url{https://doi.org/10.1117/12.2254234}
\BIBentrySTDinterwordspacing

\end{thebibliography}

\end{document}